\documentclass[slac_one]{revtex4}
\usepackage{graphicx}
\usepackage{fancyhdr}
\begin{document}

\title{Multi-Loop Results, Charm- and Bottom-Quark Masses\\and the Strong Coupling Constant}
\author{J.\,H.~K\"uhn}
\affiliation{Institut f\"ur Theoretische Teilchenphysik, Universit\"at Karlsruhe, D-76128 Karlsruhe, Germany}

\begin{abstract}
  The impact of recent multi-loop calculations on precise determinations of charm- and bottom-quark masses and the strong coupling constant is discussed.
\end{abstract}


\maketitle

\thispagestyle{fancy}

QCD has developed from a qualitative model for strong interactions into a quantitative theory with precise predictions for a multitude of observables. This development, which has taken place in particular during the past two decades, is due to significant improvements in our theoretical understanding, in calculational techniques and in improved experimental results, ranging from low energy studies in $\tau$-lepton decays through precise cross section measurements in electron-positron annihilation up to measurements of the $Z$-boson decay rate into hadrons with permille accuracy. Many of these inclusive observables are related to essentially the same object in quantum field theory, namely the absorptive part of the current-current correlator. Restricting the discussion to measurements at high energies or to properly chosen integrals, these observables can be evaluated in perturbation theory and lead to precise determinations of the fundamental parameters of QCD, quark masses and the strong coupling constant.

\section{Quark masses}

A detailed analysis of $m_c$ and $m_b$ based on the ITEP sum rules \cite{Novikov:1977dq} has been performed several years ago \cite{Kuhn:2001dm}. During the past years new and more precise data for $\sigma (e^+e^-\to\mbox{hadrons})$ have become available in the low energy region, in particular for the parameters of the charmonium and bottomonium resonances. Furthermore, the error in the strong coupling constant $\alpha_s(M_Z)$ which enters this analysis has been reduced. Last not least, the vacuum polarization induced by massive quarks has been computed in four-loop approximation; more precisely: its first derivative at $q^2=0$, which corresponds to the lowest moment of the familiar $R$-ratio has been evaluated in \cite{Chetyrkin:2006xg,Boughezal:2006px}. Based on these developments a new determination of the quark masses has been performed in Ref.~\cite{Kuhn:2007vp}. More recently also the second moment has been calculated \cite{Marq}.

The extraction of $m_Q$ from low moments of the cross section $\sigma(e^+e^-\to Q\bar{Q})$ exploits its sharp rise close to the threshold for open charm and bottom production. By evaluating the moments
\begin{eqnarray}
  {\cal M}_n^{\rm exp} &\equiv& \int \frac{{\rm d}s}{s^{n+1}} R_Q(s) \ , \label{eq::Mexp}
\end{eqnarray}
with low values of $n$, the long distance contributions are averaged out and ${\cal M}_n$ involves short distance physics only, with a characteristic scale of order $E_{\rm threshold}=2 m_Q$. Through dispersion relations the moments are directly related to derivatives of the vacuum polarization function at $q^2=0$,
\begin{eqnarray}
  {\cal M}_n^{\rm theor} = \frac{12\pi^2}{n!}\left(\frac{{\rm d}}{{\rm d}q^2}\right)^n\Pi_Q(q^2)\Bigg|_{q^2=0} \equiv \frac{9}{4}Q_Q^2\left(\frac{1}{4m_Q^2}\right)^n\bar{C}_n \ , \label{eq::Mexp1}
\end{eqnarray}
which can be evaluated in perturbative QCD ($m_Q=m_Q(\mu)$ is the $\overline{\rm MS}$ mass at the scale $\mu$). The perturbative series for the coefficients $\bar{C}_n$ in order $\alpha_s^2$ was originally evaluated up to $n=8$ in Ref.~\cite{Chetyrkin:1995ii}, and to ``arbitrary'' high order in \cite{Boughezal:2006uu,Maier:2007yn}. The four-loop contributions to $\bar{C}_0$ and $\bar{C}_1$ were evaluated in Refs.~\cite{Chetyrkin:2006xg,Boughezal:2006px}, those to $\bar{C}_2$ in \cite{Marq}. Combining Eqs. (\ref{eq::Mexp}) and (\ref{eq::Mexp1}) the quark mass can be extracted. At this point it should be emphasized that the relative weight of resonances and continuum is quite different in the experimental moments. Furthermore, low moments are less sensitive to non-perturbative contributions from condensates, to the Coulombic higher order effects, the variation of $\mu$ and the parametric $\alpha_s$ dependence. For $n=1$:
\begin{eqnarray}
  m_c(3~\mbox{GeV}) &=& 0.986(13)~\mbox{GeV} \ . \label{eq::mc3final}
\end{eqnarray}
The moment with $n=2$ is less sensitive to data for $R(s)$ from the continuum region above $5\,\mbox{GeV}$, where experimental results are scarce and the aforementioned theory uncertainties are still relatively small. The agreement between $n=1$ and $n=2$ ($m_c(3\,\mbox{GeV})=0.976(16)\,\mbox{GeV}$), together with the nice convergence with increasing order in $\alpha_s$ can be considered as additional confirmation of this approach.

Instead of measuring the moments ${\cal M}_n^{\rm exp}$ in $e^+e^-$ annihilation they can also be determined in lattice simulations. This approach has recently been pioneered in \cite{Allison:2008xk} using the Highly Improved Staggered Quarks (HISQ) discretization of the quark action in combination with four-loop perturbative results \cite{Chetyrkin:2006xg,Boughezal:2006px,Kuhn:2007vp,Sturm:2008eb,Marq}. The final result, $m_c(3~\mbox{GeV}) = 0.986(10)~\mbox{GeV}$ corresponds to a scale-invariant mass $m_c(m_c)=1.268(9)\,\mbox{GeV}$ and is in excellent agreement with the determinations based on $e^+e^-$ data.

\begin{figure}[t]
  \begin{center}
     \leavevmode\includegraphics[width=18pc]{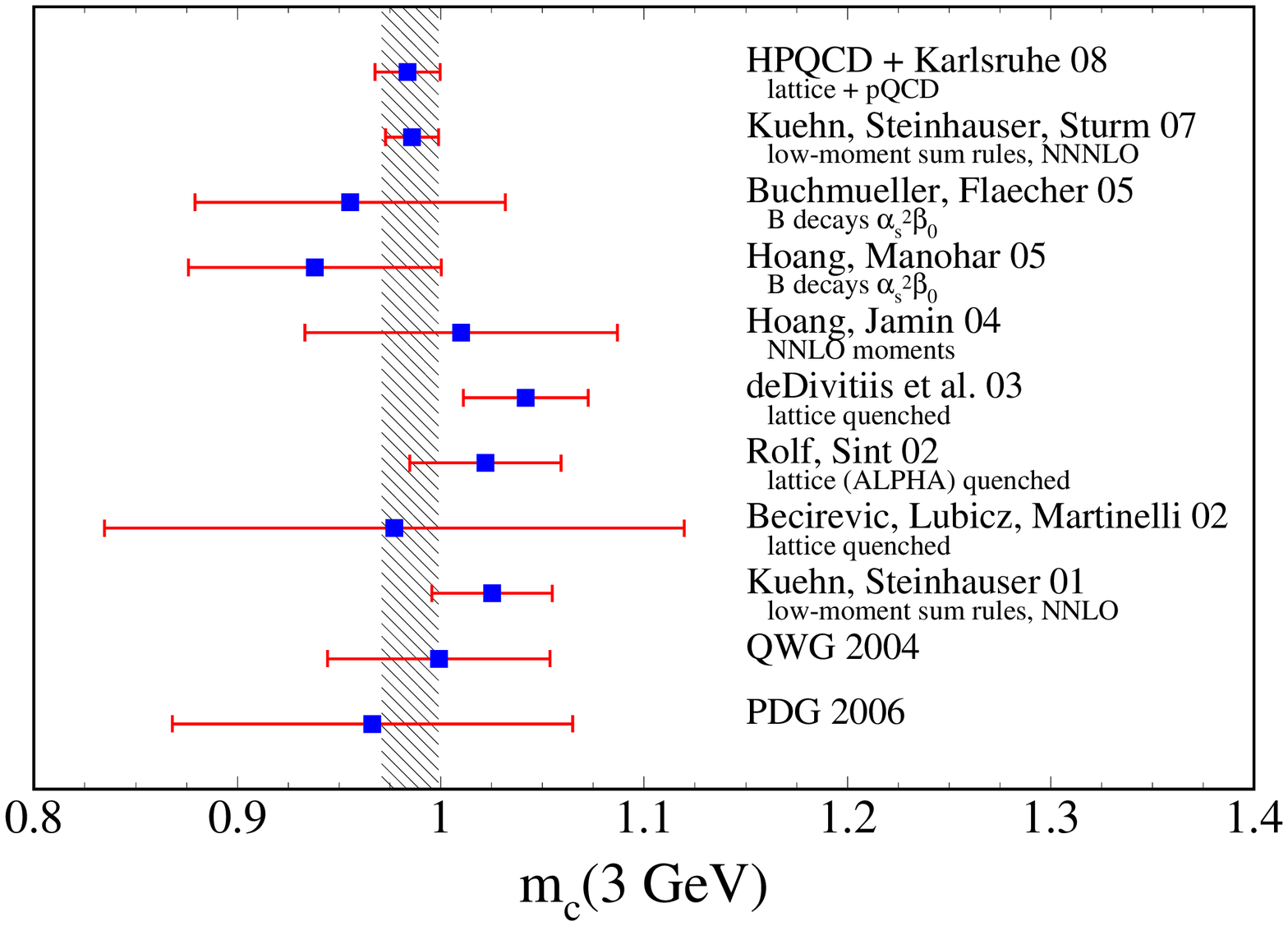}\includegraphics[width=18pc]{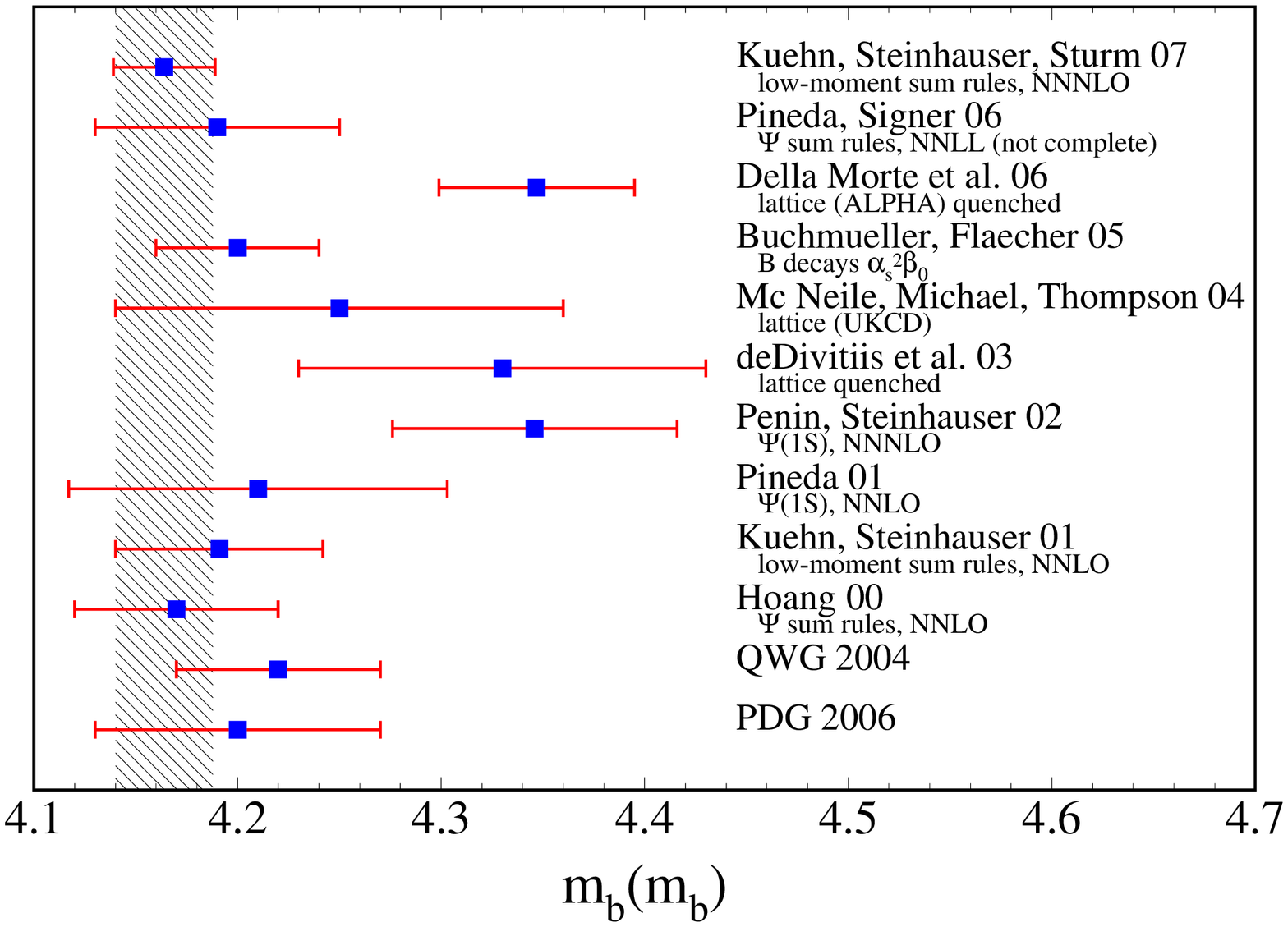}
  \end{center}
  \vspace*{-2em}
  \caption{Comparison of recent determinations of $m_c(3\,\mbox{GeV})$ and $m_b(m_b)$.\label{fig::mb_compare}}
\end{figure}

The approach based on $e^+e^-$ data is also applicable to the determination of $m_b$. The three results based on $n=1,2$ and 3 are of comparable precision. The relative size of the contributions from the threshold and the continuum region decreases for the moments $n=2$ and 3. On the other hand, the theory uncertainty is still small. Therefore the result from $n=2$ was taken as the final answer \cite{Kuhn:2007vp}, despite the fact that $\bar{C}_2$ was not yet known. The result, $m_b(10~{\rm GeV} ) = 3.609(25)~\mbox{GeV}$, corresponds to  $m_b(m_b) = 4.164(25)~\mbox{GeV}$. The recent evaluation \cite{Marq} of $\bar{C}_2$ has lead to a decrease of the central value by $2~\mbox{MeV}$ and a reduction of the error from $25~\mbox{MeV}$ to $19~\mbox{MeV}$. A comparison of a few selected $m_b$- and $m_c$-determinations is shown in Fig.~\ref{fig::mb_compare}.

\section{The strong coupling constant}

One of the most precise and theoretically safe determination of $\alpha_s$ is based on measurements of the cross section for electron-positron annihilation into hadrons \cite{ChKK:Report:1996}. These have been performed in the low-energy region between 2~GeV and 10~GeV and, in particular, at and around the $Z$ resonance at 91.2~GeV. Conceptually closely related is the measurement of the semileptonic decay rate of the $\tau$-lepton, leading to a determination of $\alpha_s$ at a scale below 2 GeV \cite{Davier:2005xq}. The perturbative expansion for the ratio $R(s)\equiv \sigma(e^+e^-\to {\rm hadrons}) / \sigma(e^+e^-\to\mu^+\mu^-)$ in numerical form is given by
\begin{eqnarray}
  R &=& 1  + a_s +  (1.9857 - 0.1152\, n_f)\, a_s^2 + (-6.63694 - 1.20013 n_f - 0.00518 n_f^2 ) \, a_s^3 \label{R_numerical_nl}\nonumber\\
    &+&(-156.61 + 18.77\, n_f - 0.7974\, n_f^2 + 0.0215\,  n_f^3 ) \, a_s^4 \ .
\end{eqnarray}
Here $a_s \equiv \alpha_s/\pi$ and the normalization scale $\mu^2=s$. The $a_s^4$ corrections are conveniently classified according to their power of $n_f$, with $n_f$ denoting the number of light quarks. The $a_s^4 n_f^3$ term is part of the ``renormalon chain'', the next term of order $a_s^4 n_f^2$ was evaluated in \cite{ChBK:vv:as4nf2}, the complete five-loop calculation has been performed in \cite{PRL}.

Let us now move to the analysis of present data for $e^+e^-$ annihilation and $\tau$ decays. Measurements of $R(s)$ at lower energies, with their correspondingly larger values of $\alpha_s$, are in principle more sensitive to $\alpha_s(M_Z)$ if the same relative precision could be obtained. At present, however, the systematic experimental error of 2\% is a limiting element for a competitive measurement. The final result of a recent analysis \cite{Kuhn:2007tc} $\alpha_s^{(4)}(9\,\mbox{GeV})=0.182 ^{+0.022}_{-0.025}$ represents the combined information on the strong coupling from $R$ measurements in the region between $3\,\mbox{GeV}$ and the bottom threshold. In contrast to the situation below 10 GeV the extraction of $\alpha_s$ from $Z$-decays is affected by the $\alpha_s^4$ terms. The ${\cal O}(\alpha_s^3)$ analysis of the electroweak working group \cite{Alcaraz:2007ri} is shifted \cite{PRL} by $\delta \alpha_s(M_Z) = 0.0005$ and one finds
\begin{equation}
  \alpha_s(M_Z)^{N\!N\!N\!L\!O} =  0.1190 \pm {0.0026}^{\rm exp} \ . \label{alpha_MZ_new}
\end{equation}
The theory  error can now safely be neglected.

Higher orders are of particular relevance in the low-energy region, for example in $\tau$ decays. The correction from perturbative QCD to the ratio $\Gamma(\tau\rightarrow {\rm hadrons}_{S=0}+\nu_\tau)/\Gamma(\tau\rightarrow l+\bar{\nu}_l+\nu_\tau)$ is given by the factor
\begin{eqnarray}
  1+\delta_0=2\int_0^{M_{\tau}^2}\frac{ds}{M_{\tau}^2}\left(1-\frac{s}{M_{\tau}^2}\right)^2\left(1+\frac{2s}{M_\tau^2}\right)\,R(s) \ , \nonumber \label{equivalent.repr}
\end{eqnarray}
which can be evaluated in Fixed Order perturbation theory or with ``Contour Improvement'' as proposed in \cite{Pivovarov:1991rh,LeDiberder:1992te}:
\begin{equation}
  \delta_0^{FO} = a_s + 5.202\,a_s^2 + 26.366\,a_s^3 + 127.079\,a_s^4 \ , \qquad \delta_0^{CI} = 1.364\,a_s + 2.54\,a_s^2 + 9.71\,a_s^3 + 64.29\,a_s^4 \ .
\end{equation}
Starting from $\delta_0^{\rm exp}= 0.1998 \pm 0.0043_{\rm exp} \label{delta_0_exp}$ \cite{Davier:2007ym} one obtains \cite{PRL} $\alpha_s^{FO}(M_{\tau}) = 0.322 \pm 0.004\relax{_{\rm exp}} \pm  0.02$ and $\alpha_s^{CI}(M_{\tau}) = 0.342 \pm 0.005\relax{_{\rm exp}} \pm  0.01$, respectively.
The second uncertainty corresponds to a change in the renormalization scale $\mu^2/M_{\tau}^2$ between $0.4$ and $2$.

As stated above the theory error for $\alpha_s$ from $Z$ decays is small compared to the experimental uncertainties. For $\tau$ decays the difference between FORT and CIPT must be considered as irreducible uncertainty \cite{PRL}:
\begin{equation}
  \alpha_s(M_\tau) =0.332 \pm 0.005_{\rm exp} \pm 0.015_{\rm th} \ . \label{als_tau}
\end{equation}
Applying four-loop running and matching \cite{vanRitbergen:1997va,Czakon:2004bu,Chetyrkin:2005ia,Schroder:2005hy} (with negligible error from the evolution from $M_\tau$ to $M_Z$):
\begin{equation}
  a_s(M_Z) = 0.1202 \pm 0.0006_{\rm exp} \pm 0.0018_{\rm th} = 0.1202 \pm 0.0019 \label{eq:asres_mz} \ .
\end{equation}

The shifts in $\alpha_s(M_Z)$ from Z- and $\tau$-decays, are opposite in sign and move  the values in the  {\em proper} direction, decreasing, thus, the current slight mismatch between two independent  determinations of $\alpha_s$. The two results are in remarkable agreement and can be combined to one of the most precise and presently only $N^3LO$ result:
\begin{equation}
  \alpha_s(M_Z) =0.1198 \pm  0.0015 \ .
\end{equation}

As discussed above, $\alpha_s$ from $\tau$-decays is strongly affected by theory uncertainties. This is reflected in three recent publications which all are based on the same set of $\tau$ data and the new five-loop results, which, however, arrive at significantly different estimates of higher order terms and hence of $\alpha_s$. In \cite{Davier:2008sk} it is argued that FOPT exhibits a poorly ``convergent'' series, in contrast to CIPT, where a subset of higher order contributions (so-called $\pi^2$-terms) is automatically summed. This analysis leads to $\alpha_s(M_\tau)=0.344\pm0.005_{exp}\pm0.007_{th}$. The small theory uncertainty is a consequence of the (artificially) reduced $\mu$-dependence and the restrictive assumptions about higher orders in perturbation theory. The opposite viewpoint has been advocated in \cite{Beneke:2008ad} where an explicit and plausible behaviour of the perturbative series is modelled. Strong cancellations are observed between the aforementioned $\pi^2$-terms and those terms which can only be obtained from a concrete higher order calculation.

All these papers are based on similar assumptions about non-perturbative power-suppressed contributions. These assumptions have been questioned in \cite{Maltman:2008nf}, employing different weight functions in order to suppress these poorly determined terms with dimension $D>8$. A significantly smaller result, $\alpha_s(M_\tau)=0.3209\pm0.0046\pm0.0018$, is obtained within CIPT. In total the spread among the different results is covered reasonably well by eq. (\ref{als_tau}).

Last not least it is instructive to compare eqs. (\ref{alpha_MZ_new}) and (\ref{eq:asres_mz}) with $\alpha_s$ determinations based on lattice simulations. The most recent results from a precision simultaion by the HPQCD collaboration \cite{Davies:2008sw} $\alpha_s(M_Z)=0.1183\pm0.0007$ and from a study \cite{Maltman:2008bx} based on essentially the same lattice data and an alternative perturbative analysis, $\alpha_s(M_Z)=0.1192\pm0.0011$ are in mutual agreement, as should be expected because they are based on a similar set of data.

Remarkably enough, a completely different approach \cite{Allison:2008xk}, which exploits the lattice simulation of the pseudoscalar correlator and which has been mentioned already above leads to $\alpha_s(M_Z)=0.1174\pm0.0012$ again well consistent with the other determinations.

\begin{acknowledgments}
  This work was supported by SFB/TR-9 ``Computational Particle Physics''.
\end{acknowledgments}

\end{document}